\documentclass[aps,prl,twocolumn,showpacs,superscriptaddress,floatfix,nofootinbib]{revtex4}

\usepackage{graphicx}
\usepackage{epsfig}
\usepackage[english]{babel}

\newcommand{\be}{\begin{equation}}
\newcommand{\bea}{\begin{eqnarray}}
\newcommand{\eea}{\end{eqnarray}}
\newcommand{\ee}{\end{equation}}
\newcommand{\bra}[1]{\mbox{$\langle #1 |$}}
\newcommand{\ket}[1]{\mbox{$| #1 \rangle$}}

\newcommand{\proj}[1]{\mbox{$|#1\rangle \!\langle #1 |$}}
\newcommand{\ev}[1]{\mbox{$\langle #1 \rangle$}}

\def\B{{\cal B}}
\def\tr{ \mbox{tr}}

\begin{document}

\title{Entanglement in quantum critical phenomena}

\author{G. Vidal}
\affiliation{Institute for Quantum Information, California Institute for Technology, Pasadena, CA 91125 USA}
\author{J. I. Latorre}
\affiliation{Dept. d'Estructura i Constituents de la Mat\` eria, Univ. Barcelona, 08028, Barcelona, Spain.}
\author{E. Rico}
\affiliation{Dept. d'Estructura i Constituents de la Mat\` eria, Univ. Barcelona, 08028, Barcelona, Spain.}
\author{A. Kitaev}
\affiliation{Institute for Quantum Information, California Institute for Technology, Pasadena, CA 91125 USA}

\date{\today}

\begin{abstract}

Quantum phase transitions occur at zero temperature and involve the appearance of long-range correlations. These correlations are not due to thermal fluctuations but to the intricate structure of a strongly entangled ground state of the system. 
We present a microscopic computation of the scaling properties of the ground-state entanglement in several 1D spin chain models both near and at the quantum critical regimes. 
We quantify entanglement by using the entropy of the ground state when the system is traced down to $L$ spins. This entropy is seen to scale logarithmically with $L$, with a coefficient that corresponds to the central charge associated to the conformal theory that describes the universal properties of the quantum phase transition. 
Thus we show that entanglement, a key
concept of quantum information science, obeys universal scaling laws as dictated by the representations
of the conformal group and its classification motivated by
string theory. 
This connection unveils a monotonicity law for ground-state entanglement along the renormalization group flow. We also identify a majorization rule possibly associated to conformal invariance and apply the present results to interpret the breakdown of density matrix renormalization group techniques near a critical point.

\end{abstract}


\maketitle


The study of entanglement in composite systems is one of the major goals of quantum information science \cite{BeDi00,book}, where entangled states are regarded as a valuable resource for processing information in novel ways. For instance, the entanglement between systems $A$ and $B$ in a joint pure state $\ket{\Psi_{AB}}$ can be used, together with a classical channel, to {\em teleport} or send quantum information \cite{Be93}. From this resource-oriented perspective, the {\em entropy of entanglement} $E(\Psi_{AB})$ measures the entanglement contained in $\ket{\Psi_{AB}}$ \cite{BBPS96}. It is defined as the von Neumann entropy of the reduced density matrix $\rho_A$ (equivalently $\rho_B$), 
\be
E(\Psi_{AB}) = -\tr \left( \rho_A \log_2 \rho_A \right),
\label{eq:ent}
\ee
and directly determines, among other aspects, how much quantum information can be teleported by using $\ket{\Psi_{AB}}$.


On the other hand, entanglement is appointed to play a central role in the study of strongly correlated quantum systems \cite{Pr99,OsNi01,ZaWa02}, since a highly entangled ground state is at the heart of a large variety of collective quantum phenomena. Milestone examples are the entangled ground states used to explain superconductivity and the fractional quantum Hall effect, namely the BCS ansatz \cite{BCS57} and the Laughlin ansatz \cite{La83}. Ground-state entanglement is, most promisingly, also a key factor to understand quantum phase transitions \cite{Sa99,OsNi02}, where it is directly responsible for the appearance of long-range correlations. Consequently, a gain of insight into phenomena including, among others, Mott insulator-superfluid transitions, quantum magnet-paramagnet transitions and phase transitions in a Fermi liquid is expected by studying the structure of entanglement in the corresponding underlying ground states.


In the following we analyze the ground-state entanglement near and at a quantum critical point in a series of 1D spin-$1/2$ chain models. In particular, we consider the hamiltonians
\be
H_{XY} = -\sum_{l=0}^{N-1} \left( \frac{a}{2}[(1+\gamma) \sigma_l^x \sigma_{l+1}^x + (1-\gamma) \sigma_l^y \sigma_{l+1}^y] + \sigma_l^z  \right)
\label{eq:XY}
\ee
and
\be
H_{XXZ} =  -\sum_{l=0}^{N-1} \left(\sigma_l^x \sigma_{l+1}^x + \sigma_l^y \sigma_{l+1}^y + \Delta \sigma_l^z \sigma_{l+1}^z + \lambda \sigma_l^z \right),
\label{eq:XXZ}
\ee
which contain both first-neighbor interactions and an external magnetic field, and are used to describe a range of $1D$ quantum systems \cite{Sa99}. Notice that $H_{XY}(\gamma = 1)$ corresponds to the Ising chain, whereas $H_{XXZ}(\Delta=1)$ describes spins with isotropic Heisenberg interaction. Both Hamiltonians coincide for $\gamma=\Delta=0$, where they become the XX model. 


Osterloh {\em et al} \cite{Os02} and Osborne and Nielsen \cite{OsNi02} have recently considered the entanglement in the XY spin model, Eq. (\ref{eq:XY}), in the neighborhood of a quantum phase transition. Their analysis, focused on {\em single-spin} entropies \cite{OsNi02} and on {\em two-spin} quantum correlations \cite{Os02,OsNi02}, suggestively shows that these one- and two-spin entanglement measures are picked either near or at the critical point.
Here, alternatively, we undertake the study of the {\em entanglement between a block of $L$ contiguous spins and the rest of the chain}, when the spin chain is in its ground state $\ket{\Psi_g}$. Thus, the aim in the following is to compute the entropy of entanglement, Eq. (\ref{eq:ent}), for the state $\ket{\Psi_g}$ according to bipartite partitions parameterized by $L$,
\be
S_L \equiv - \tr \left( \rho_{L} \log_2 \rho_{L} \right),
\label{eq:SL}
\ee
where $\rho_L \equiv \tr_{\bar{\B}_L} \proj{\Psi_g}$ is the reduced density matrix for $\B_L$, a block of $L$ spins.

The XY model, Eq. (\ref{eq:XY}), is an {\em exactly solvable} model, in that $H_{XY}$ can be diagonalized by first using a Jordan-Wigner transformation into fermionic modes and by subsequently concatenating a Fourier transformation and a Bogoliubov transformation (see for instance \cite{Sa99}). The calculation of $S_L$, as sketched below, also uses the fact that the ground state $\ket{\Psi_g}$ of $H_{XY}$ and the corresponding density matrices $\rho_L$ are all {\em gaussian} states that can be completely characterized by means of certain correlation matrix of second moments. 

More specifically, let us introduce two Majorana operators, $c_{2l}$ and $c_{2l+1}$, on each site $l = 0, \cdots, N-1$ of the spin chain,
\be
c_{2l} \equiv \left( \prod_{m=0}^{l-1} \sigma_m^z \right) \sigma_l^x; ~~~
c_{2l+1} \equiv \left( \prod_{m=0}^{l-1} \sigma_m^z \right) \sigma_l^y.
\label{eq:cc}
\ee
Operators $c_m$ are hermitian and obey the anti-commutation relations $\{c_m, c_n\}= 2\delta_{mn}$. Hamiltonian $H_{XY}$ can be rewritten as
\begin{eqnarray*}
H_{XY} = i\sum_{l=0}^{N-1} \left(~ \frac{a}{2}~[~(1+\gamma)~c_{2l+1}c_{2l+2}
-\right.  \\
 \left.  (1-\gamma)~c_{2l}c_{2l+3}~] + c_{2l}c_{2l+1}\frac{}{}~\right),
\label{eq:majXY}
\end{eqnarray*}
and can be subsequently diagonalized by canonically transforming the operators $c_m$. The expectation value of $c_m$ when the system is in the ground state, i.e. $\ev{c_m}\equiv \bra{\Psi_g}c_m\ket{\Psi_g}$, vanishes for all $m$ due to the ${\cal Z}_2$ symmetry $(\sigma_l^x,\sigma_l^y,\sigma_l^z)\rightarrow (-\sigma_l^x,-\sigma_l^y,\sigma_l^z)~ \forall l$ of $H_{XY}$. In turn, the expectation values
\be
\ev{c_mc_n} = \delta_{mn}+ iB_{mn},
\label{eq:corr}
\ee
completely characterize $\ket{\Psi_g}$, for any other expectation value can be expressed, by using Wick's theorem, in terms of $\ev{c_mc_n}$. Matrix B reads \cite{expanded}
\be
B = \left[
 \begin{array}{ccccc}
\Pi_0  & \Pi_1   &    \cdots & \Pi_{N-1}  \\
\Pi_{-1} & \Pi_0   & &\vdots\\
\vdots&  & \ddots&\vdots  \\
\Pi_{1-N} & \cdots & \cdots  &  \Pi_0 
\end{array}
\right], ~~~ \Pi_l = \left[\begin{array}{cc}
0 & g_l \\
-g_{-l} & 0
\end{array}
\right],
\label{eq:B}
\ee
with real coefficients $g_l$ given, when $N\rightarrow \infty$, by 
\be
g_l = \frac{1}{2\pi}\int_{0}^{2\pi} d\phi e^{-il\phi}\frac{a \cos \phi - 1 - ia \gamma \sin \phi}{|a \cos \phi - 1 - ia\gamma \sin \phi|}.
\label{eq:g}
\ee

From Eqs. (\ref{eq:B})-(\ref{eq:g}) we can extract the entropy $S_L$ of Eq. (\ref{eq:SL}) as follows. First we compute the correlation matrix of the state $\rho_L$ for block $\B_L$, namely $\delta_{mn} + i(B_L)_{mn}$, where 
\be
B_L = \left[
 \begin{array}{cccc}
\Pi_0  & \Pi_1   & \cdots & \Pi_{L-1}   \\
\Pi_{-1} & \Pi_0   &  & \vdots\\
\vdots & & \ddots & \vdots\\
 \Pi_{1-L}   & \cdots &\cdots & \Pi_0 
\end{array}
\right]
\label{eq:BL}
\ee
is constructed by eliminating $2(N-L)$ contiguous columns and rows from $B$, those corresponding to the $N-L$ traced-out spins. Let $V\in SO(2L)$ denote an orthogonal matrix that brings $B_L$ into a block-diagonal form, that is
\be
\tilde{B}_L = VB_LV^T = \oplus_{m=0}^{L-1} \nu_m  \left[\begin{array}{cc}
0 & 1 \\
-1 & 0
\end{array}
\right],~~~\nu_m \geq 0.
\ee
Then the set of $2L$ Majorana operators $d_m \equiv \sum_{n=0}^{2L-1} V_{mn} c_n$, obeying $\{d_m, d_n\}= 2\delta_{mn}$, have a block-diagonal correlation matrix $\ev{d_md_n} = \delta_{mn}+i(\tilde{B}_L)_{mn}$. Therefore, the $L$ fermionic operators $b_l \equiv (d_{2l}+id_{2l+1})/2$, obeying $\{b_m,b_n\}=0$ and $\{b_m^{\dagger}b_n\}=\delta_{mn}$, have expectation values
\be
\ev{b_m}=0,~~~\ev{b_mb_n}=0, ~~~~\ev{b^{\dagger}_mb_n} = \delta_{mn}\frac{1+\nu_m}{2}.
\label{eq:correfermi}
\ee
Eq. (\ref{eq:correfermi}) indicates that the above fermionic modes are in a {\em product} or {\em uncorrelated} state, that is
\be
\rho_L=\varrho_0 \otimes \cdots \otimes \varrho_{L-1},
\ee
where $\varrho_m$ denotes the mixed state of mode $m$. The entropy of $\rho_L$ is a sum over the entropy $H_2((1+\nu_m)/2)$ of each mode [$H_2(x)= -x\log_2 x -(1-x)\log_2(1-x)$ is the binary entropy] and thus reads
\be
S_L = \sum_{m=0}^{L-1} H_2(\frac{1+\nu_m}{2}).
\label{eq:entronu}
\ee

For arbitrary values of $(a,\gamma)$ in $H_{XY}$ and in the thermodynamic limit, $N\rightarrow \infty$, one can evaluate Eq. (\ref{eq:g}) numerically, diagonalize $B_L$ in Eq. (\ref{eq:BL}) to obtain $\nu_m$ and then evaluate Eq. (\ref{eq:entronu}). The computational effort grows only polynomically with $L$ and produces reliable values of $S_L$ for blocks with up to several tens of spins. However, further analytical characterization is possible in some cases, which speeds the computation significantly. For instance, for the $XX$ model with magnetic field, $\gamma=0$ and $1/a\in [-1,1]$, one obtains $g_0=\phi_a/\pi-2$ and $g_l=2\sin(l\phi_a)/l\pi$ for $l\neq 0$, where $\phi_a = \arccos (1/a)$, from which $S_L$ can be numerically determined. We obtain, up to $L = 100$ spins,
\be
S_L^{XX} \approx \frac{1}{3} \log_2(L) + k_1(a), 
\label{eq:xxent}
\ee
where $k_1(a)$ depends only on $a$. For the Ising model with critical magnetic field, $\gamma=1, a=1$, one finds $g_l = -2/l$ for odd $l$ and $g_l=0$ for even $l$ and, for up to $L=100$ spins,
\be
S_L^{Ising} \approx \frac{1}{6} \log_2(L) + k_2, ~~~~~k_2 \approx \pi/3.
\label{eq:isingent}
\ee
Finally, for the Ising model with magnetic field, $\gamma=1$, and for $a$ close to 1, Kitaev has obtained an analytical expression for the entropy of half of an infinite chain \cite{Ki02}. To use this result in our setting we need to double its value because the entropy resides near the boundary. Thus we get
\be
S_{N/2}^{Ising} \approx \frac{1}{6} \log_2 \frac{1}{|1-a|}. 
\label{eq:isingKi}
\ee

The XXZ model, Eq. (\ref{eq:XXZ}), cannot be analyzed using the
previous method. Instead we have used the Bethe ansatz
\cite{Be} to exactly determine, through a numerical procedure, the
ground state $\ket{\Psi_g^{(20)}}$ of $H_{XXZ}$ for a chain of up to
$N=20$ spins, from which $S_L^{(20)}$ can be computed. Recall that in
the XXZ model the non-analyticity of the ground-state energy
characterizing a phase transition occurs already for a finite number
$N$ of spins in the chain, since it is due to level-crossing. It turns
out that, correspondingly, already for $N=20$ spins one can observe a
distinct, characteristic behavior of $S^{(20)}_L$ depending on whether
the values $(\Delta,\lambda)$ in Eq. (\ref{eq:XXZ}) belong to a
critical regime.

The results of the computation of $S_L$ for the spin chains (\ref{eq:XY}) and (\ref{eq:XXZ}) can be summarized as follows.

\begin{figure}
\epsfig{file=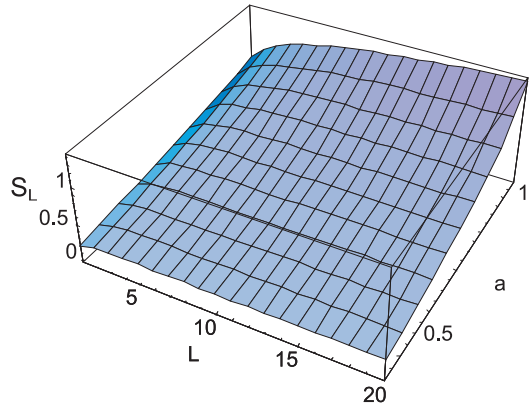,width=8.8cm}
\caption{\label{fig:ising} Non-critical $S_L$ for the Ising model, $H_{XY}(\gamma=1)$ in Eq. (\ref{eq:XY}), as
a function of the size $L$ of the spin block (left axis) and parameter
$a$ (right axis). The finite correlation length governing non-critical phenomena translates into a finite {\em entanglement length}, that is, a finite value
of $L$ for which adding new spins to a block does not increase its
entanglement with the rest of the chain. Such an {\em entanglement
length} (which scales as the correlation length) diverges only at the critical point $a=1$. Scaling arguments also imply that the entropy surface is given by $\log[Lf(L|1-a|)]$. For any given $a$, the saturation value for the entropy is given by Eq. (\ref{eq:isingKi}).
}
\end{figure}

{\em Non-critical regime.} For those values $(a,\gamma)$ or $(\Delta,\lambda)$ for which the models are non-critical, the entropy of entanglement $S_L$ either vanishes for all $L$ [{\em e.g.} when a sufficiently strong magnetic field aligns all spins into a product, unentangled state] or grows monotonically as a function of $L$ until it reaches a {\em saturation} value $S_{max}$.
For instance, in the infinite Ising chain the saturation entropy $S_{max}$ is given by Eq. (\ref{eq:isingKi}). As shown in Fig. (\ref{fig:ising}), $S_L$ often approaches $S_{max}$ already for a small number $L$ of spins.

\begin{figure}
\epsfig{file=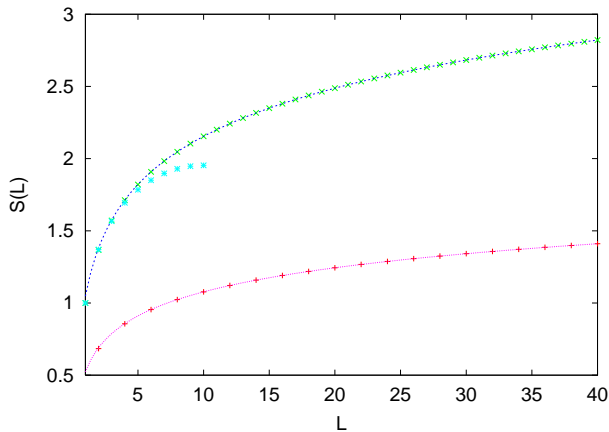,width=8.4cm}
\caption{\label{fig:all} Numerical calculation of critical
entanglement $S_L$ for 1D spin chains. Points ($+$) come from an
infinite Ising chain with critical magnetic field,
$H_{XY}(a=1,\gamma=1)$ in Eq. (\ref{eq:XY}), and corresponds to
$S_L^{Ising} \approx \log_2(L)/6+k_2$.  Curve ($\times$) comes from an
infinite XX spin chain without magnetic field, $H_{XY}(a=\infty,
\gamma=0)$ in Eq. (\ref{eq:XY}), and corresponds to $S^{XX}_L \approx
\log_2(L)/3+k_1$ . Thus the growth in $L$ of entanglement in the XX model, asymptotically described by a
free boson, is twice that in the Ising model,
corresponding to a free fermion,  $S^{XX}_{L+1}-S^{XX}_L \approx 2(S_{L+1}^{Ising}-S_L^{Ising})$. 
Finally, curve ($*$) comes from a XXX chain of $20$ spins
without magnetic field, $H_{XXZ}(\Delta=1, \lambda=0)$ in
Eq. (\ref{eq:XXZ}). These finite-chain results, obtained using the
Bethe ansatz, combine the logarithmic behavior ($L=1,\cdots, 5$) of a
free boson field theory with a finite-size saturation effect
($L=6,\cdots, 10$). We have added the lines ${c+\bar c\over 6} [\log_2(L)+\pi]$ both for bosons and fermions to highlight their remarkable
agreement with the numerical diagonalization.}
\end{figure}

{\em Critical regime.} Instead, critical ground-states are characterized by an entropy $S_L$ that diverges logarithmically with $L$,
\be
S_L \approx \frac{c+\bar{c}}{6} \log_2 (L) + k,
\label{eq:critent}
\ee
with a coefficient given by the holomorphic and antiholomorphic {\em central charges} $c$ and $\bar{c}$ of the conformal field theory that describes the universal properties of the phase transition \cite{cft}, see Fig. (\ref{fig:all}). This expression was derived by Holzhey, Larsen and Wilczek \cite{Ho94} for the {\em geometric entropy} (analogous of Eq. (\ref{eq:SL}) for a conformal field theory), and our calculation confirms it for several critical spin chains.  Thus, the critical Ising model corresponds to a free fermionic field theory, with central charges $c_f=\bar{c}_f=1/2$, whereas the rest of critical regimes in (\ref{eq:XY}) and (\ref{eq:XXZ}) are described by a free bosonic field theory, $c_b=\bar{c}_b=1$ (cf. Eqs. (\ref{eq:xxent}) and (\ref{eq:isingent})). In particular, the marginal deformation $0<\gamma \leq 1$ for $a=1$ shows scaling for
every $\gamma$ with universal coefficient $c=\bar{c}=1/2$.  It is possible to compare the subleading correction between two
different values of $\gamma$. The behavior we obtain is described by 
\be 
\lim_{L\rightarrow \infty} \left[ S_L (\gamma=1)-S_L(\gamma) \right] = -{c+\bar c\over 6} \log_2 \gamma \ .
\ee
The singular behavior at $\gamma=0$ is the signature of the fact that that
point belongs to the abrupt change of universality class for the XX model.

The above characterizations motivate a number of observations, that we move to discuss.

Critical and non-critical ground states contain structurally different forms of quantum correlations. Non-critical ground-state entanglement corresponds to a weak, {\em semi-local} form of entanglement driven by the appearance of a length scale, {\em e.g.} a mass gap. Indeed, for any $L$, the reduced density matrix $\rho_L$ is supported on a small, bounded subspace of the Hilbert space of the $L$ spins, and can be obtained by diagonalizing the hamiltonian corresponding to the block $\B_L$ and only a few extra {\em neighboring} spins, as skillfully exploited in White's density matrix renormalization group (DMRG) techniques \cite{Wh93}. We note here that a bounded rank for $\rho_L$ (relatedly, a saturation value for $S_L$) is instrumental for the success of DMRG schemes, where only of a finite number of eigenvectors of $\rho_L$ can be kept. Critical bound-state entanglement corresponds, on the contrary, to a stronger form of entanglement, one that embraces the system at all length scales simultaneously. DMRG techniques have reportedly failed to reproduce quantum critical behavior \cite{RO99} and we may, in view of Eq. (\ref{eq:critent}), be in a position to understand why. Indeed, the divergent character of $S_L$ is just one particular manifestation of the fact that the number of relevant eigenvectors of $\rho_L$ unboundedly grows with $L$. If, as is the case in DMRG schemes, only a finite number of levels can be considered, then a sufficiently large $L$ will always make the computation of $\rho_L$ impossible (arguably, even in an approximate sense) by using such schemes. This strongly suggests that overcoming the above difficulties necessarily requires techniques that do not attempt to reproduce the critical behavior of the ground state through a local, real space construction.

Another remarkable, far-reaching fact is that, as mentioned below
Eq. (\ref{eq:critent}), our results coincides with entropy
computations performed in conformal field theory. There, a {\em
geometric} or {\em fine-grained} entropy analogous to
Eq. (\ref{eq:SL}) but for a continuous field theory has been
considered by several authors, including Srednicki \cite{Sr93},
Callan and Wilczek \cite{Ca94},
Holzhey et al \cite{Ho94} and Fiola et al \cite{Fi94}. Thus, starting
from non-relativistic spin chain models, and by performing a
microscopic analysis of a relevant quantity in quantum information, we
have obtained a universal scaling law for entanglement that is in full
agreement with previous findings in the context of, say, black-hole
thermodynamics in 1+1 dimensions \cite{Ho94,Fi94}.

The above connection has a number of implications to be exploited. For
instance, Srednicki \cite{Sr93} has obtained the behavior of entropy
in 2+1 and 3+1 dimensional conformal field theories. For a region
${\cal R}$ in $2$ or $3$ spatial dimensions, the entropy of ${\cal R}$
is proportional to the size $\sigma({\cal R})$ of its boundary, \be
S_{\cal R} \approx \kappa\sigma({\cal R}). \label{eq:d} \ee That is,
the entropy per unit of boundary area, $\kappa$, is independent of the
size of ${\cal R}$. [This is in sharp contrast with the same quantity
in 1D, where the boundary consists of two points and $S_{\cal
R}/\sigma({\cal R})$ diverges logarithmically with the length $L$ of
${\cal R}$]. Accordingly, Eq. (\ref{eq:d}) also describes the critical
ground-state entanglement of 2D and 3D spin lattices. 

Also the fact that the entropy of entanglement for $1D$ critical spin chains,  Eq. (\ref{eq:critent}), matches well-known conformal field theory parameters carries an extra bonus. The coefficient in control of the
divergent behavior of $S_L$ at critical points is the central
charge, which is subject to Zamolodchikov's c-theorem \cite{Za}. The c-theorem
states that the central charges associated to the ultraviolet
and infrared end points
of renormalization group flows, labeled by $C_{UV}$ and
$C_{IR}$, obey the inequality
$C_{UV} > C_{IR}$ for unitary theories. This powerful result
establishes an irreversible arrow as renormalization group
transformations are performed. The translation of this idea
to the quantum information setting is that entanglement decreases
along renormalization group flows. An infrared theory carries
less global entanglement than the ultraviolet theory where
it flowed from. The c-theorem seems natural as renormalization
group transformations integrate out short distance degrees
of freedom, accompanied with their quantum correlations. Yet,
it is not at all trivial due to, first, the infinite degrees of freedom (needing regularization) existing in a quantum field theory and,
second, the rescaling
step in the renormalization group transformation.
It is noteworthy, then, that entanglement decreases both ($i$) under 
the {\em local operations and classical communication} and ($ii$) along {\em renormalization group trajectories}.
The former case corresponds to {\em local manipulation} of an entangled system while the second is made out of a block-spin
transformation followed by a {\em rescaling} of the system.
Both actions do reduce quantum correlations and become
irreversible \cite{Za,VC01}.

One more remark. From Eqs. (\ref{eq:g})-(\ref{eq:BL}) the complete spectrum of $\rho_L$ can be extracted. The $2^L$ eigenvalues are given by
\be
\lambda_{x_1x_2\cdots x_L} = \prod_{m=0}^{L-1} \frac{1+(-1)^{x_m}\nu_m}{2},~~~x_m = 0,1~ \forall m.
\ee
This allows us to look in more detail to the reshuffling of the ground state as more sites are incorporated in the block $\B_L$.
Every time a new spin is added, the amount of local surprise due to quantum correlations with the rest of the chain increases, and so does the
entropy. But critical quantum correlations entangle every single subset of the system, and the way they are reordered is far more subtle than
the relation hinted by entropy arguments. We have numerically verified that also a majorization relation \cite{Bh96} holds for the ground-state reduced density matrices of the infinite XY spin chain, namely 
\be
\vec\lambda_{L+2} \prec \vec\lambda_L \quad \forall L,
\label{eq:maj} 
\ee
where the jump in steps of two is forced by the subtleties of the microscopic
model.
Thus, a critical ground state orderly redistributes weights, so as to accommodate for the new correlations, according to a detailed, exponentially large set of inequalities as contained in Eq. (\ref{eq:maj}). In this sense, majorization  may be a signature --admittedly a very refined one-- of conformal invariance. 

But the majorization counterpart in the continuum conformal field theory is not yet known. Perhaps, then, the correspondence between concepts of quantum information science and conformal field theory, between critical ground-state entanglement and geometrical entropy, can also be exploited in the reverse direction.

This work was supported by the by the Spanish grants
GC2001SGR-00065 and MCYT FPA2001-3598, by the National Science
Foundation of USA under grant EIA--0086038,
and by the European Union under grant ISF1999-11053.


\begin{thebibliography}{99}

\bibitem{BeDi00} C. H. Bennett and D. P. DiVincenzo, Nature {\bf 404}, 247 - 255 (2000). 

\bibitem{book} M. A. Nielsen and I. L. Chuang, {\em Quantum computation and quantum communication} (Cambridge Univ. Press, Cambridge, 2000).

\bibitem{Be93} C. H. Bennett, G. Brassard, C. Cr\'epeau, R. Josza, A. Peres and W.K. Wootters, Phys. Rev. Lett. {\bf 70}, 1895 (1993).

\bibitem{BBPS96}
C. H. Bennett, H. J. Bernstein, S. Popescu, and B. Schumacher,
Phys. Rev. A {\bf 53}, 2046-2052 (1996).

\bibitem{Pr99} J. Preskill, J. Mod. Opt. {\bf 47} (2000) 127-137,

\bibitem{OsNi01} T. J. Osborne and M. A. Nielsen, {\em Entanglement, quantum phase transitions, and density matrix renormalization}, quant-ph/0109024.

\bibitem{ZaWa02} P. Zanardi and X. Wang,  
{\em Fermionic entanglement in itinerant systems}, J. Phys. A {\bf 35} (2002) 7947

\bibitem{BCS57} 
J. Bardeen, L. N. Cooper and J. R. Schrieffer, Phys. Rev. {\bf 108}, 1175 (1957).

\bibitem{La83} R. B. Laughlin, Phys. Rev. Lett. {\bf 50}, 1395 (1983).

\bibitem{Sa99} S. Sachdev, {\em Quantum phase transitions}, (Cambridge University Press, Cambridge, 1999).

\bibitem{OsNi02} T. J. Osborne and M. A. Nielsen,
{\em Entanglement in a simple quantum phase transition}, quant-ph/0202162.

\bibitem{Os02} A. Osterloh, L. Amico, G. Falci and R. Fazio, 
{\em Scaling of entanglement close to a quantum phase transition},
Nature {\bf 416} (2002) 608-610.

\bibitem{expanded} A more detailed analysis of these manipulations, that can be extracted from \cite{Ann}, will be presented somewhere else. We note that, for the sake of simplicity, we do not discuss finite $N$ corrections to Eq. (\ref{eq:B}). In the thermodynamic limit, $N\rightarrow \infty$, they do not contribute to  Eq. (\ref{eq:BL}) for finite $L$, and therefore do not affect the computation of $S_L$. 

\bibitem{Ann} E. Lieb, T. Schultz and D. Mattis, Annals of Phys. {\bf 16} 407 (1961).

\bibitem{Ki02} A. Kitaev, {\em private communication}.

\bibitem{Be} H. A. Bethe, Z. Physik {\bf 71}, 205 (1931).

\bibitem{cft} Paul Ginsparg, {\em Applied conformal field theory}, 
    Les Houches Summer School 1988, pp 1-168.

\bibitem{Ho94} C. Holzhey, F. Larsen and F. Wilczek, Nucl. Phys. B {\bf 424} (1994) 443-467.

\bibitem{Wh93} S. R. White, Phys. Rev. B {\bf 48}, 10345-10356 (1993).

\bibitem{RO99} S. Rommer and S. \"Ostlund, {\em Density matrix renormalization}, Dresden, 1998 (Springer, Berlin, 1999), pp 67-89. 

\bibitem{Sr93} M. Srednicki, Phys. Rev. Lett. {\bf 71} (1993) 666-669.

\bibitem{Ca94} C. G. Callan and F. Wilczek, Phys. Lett. {\bf B333} (1994) 55-61.

\bibitem{Fi94} T. M. Fiola, J. Preskill, A. Strominger and S. P. Trivedi, Phys. Rev. D {\bf 50} (1994) 3987-4014.

\bibitem{Za} A. B. Zamolodchikov JETP Lett. {\bf 43} (1986) 730;
A. Cappelli, D. Friedan and J. I. Latorre, Nucl. Phys. B {\bf 352} (1991) 616;
S. Forte and J. I. Latorre, Nucl. Phys. B {\bf 535} (1998) 709.

\bibitem{VC01} G. Vidal and J. I. Cirac, Phys. Rev. Lett. {\bf 86}, (2001) 5803-5806.

\bibitem{Bh96} R. Bhatia, {\em Matrix Analisis}, Graduate Texts in Mathematics vol. 169, Springer-Verlag, 1996.

\end{thebibliography}
\end{document}